
%
%
%
\magnification=1200
%
\hoffset 0.2truein   
\hsize=6.5truein     
\vsize=8.8truein     
\topskip=15pt            
\baselineskip16pt
%
%
\fontdimen1\tenrm=0.0pt  
\fontdimen2\tenrm=4.0pt  
\fontdimen3\tenrm=7.0pt  
\fontdimen4\tenrm=1.6pt  
\fontdimen5\tenrm=4.3pt  
\fontdimen6\tenrm=10.0pt 
\fontdimen7\tenrm=2.0pt  

\font\section=cmbx10 scaled\magstep2
\font\subsection=cmbx10 scaled\magstep1

\font\figuref=cmr9

\def\ssstyle{\scriptscriptstyle}
\def\fnstyle{\baselineskip8pt \font \bx cmr9 \bx}
%
%
\def\lsim{\;\raise0.3ex\hbox{$<$\kern-0.75em\raise-1.1ex\hbox{$\sim$}}\;}
\def\gsim{\;\raise0.3ex\hbox{$>$\kern-0.75em\raise-1.1ex\hbox{$\sim$}}\;}

\def\Slash#1{#1\kern-0.5em\raise.05ex\hbox{/}}
\def\slash#1{#1\kern-0.45em\raise.05ex\hbox{{$\scriptstyle /$}}}

\def\parbarit#1{#1\kern-0.8em\raise1.5ex\hbox{$\ssstyle(-)$}}
\def\parbar#1{#1\kern-1.0em\raise1.5ex\hbox{$\scriptscriptstyle(-)$}}
\def\Parbar#1{#1\kern-1.0em\raise2.1ex\hbox{$\scriptscriptstyle(-)$}}
%
%


%
%
\def\jump{\vskip 0.5truecm}

\def\ref#1{[{#1}]}

\def\dr{{\rm d}}

%
%






\newcount\refnumber
\newcount\temp
\newcount\test
\newcount\tempone
\newcount\temptwo
\newcount\tempthr
\newcount\tempfor
\newcount\tempfiv
\newcount\testone
\newcount\testtwo
\newcount\testthr
\newcount\testfor
\newcount\testfiv
\newcount\itemnumber
\newcount\totalnumber
\refnumber=0
\itemnumber=0
\def\initreference#1{\totalnumber=#1
                 \advance \totalnumber by 1
                 \loop \advance \itemnumber by 1
                       \ifnum\itemnumber<\totalnumber
                        \temp=100 \advance\temp by \itemnumber
                        \count\temp=0 \repeat}

\def\ref#1{\temp=100 \advance\temp by #1
   \ifnum\count\temp=0
    \advance\refnumber by 1  \count\temp=\refnumber \fi
   \ [\the\count\temp]}

\def\reftwo#1#2{\tempone=100 \advance\tempone by #1
   \ifnum\count\tempone=0
   \advance\refnumber by 1  \count\tempone=\refnumber \fi
   \temptwo=100 \advance\temptwo by #2
   \ifnum\count\temptwo=0
   \advance\refnumber by 1  \count\temptwo=\refnumber \fi
 \testone=\count\tempone \testtwo=\count\temptwo
 \sorttwo\testone\testtwo
     \ [\the\testone,\the\testtwo]}       

\def\refthree#1#2#3{\tempone=100 \advance\tempone by #1
   \ifnum\count\tempone=0
    \advance\refnumber by 1  \count\tempone=\refnumber \fi
    \temptwo=100 \advance\temptwo by #2
   \ifnum\count\temptwo=0
    \advance\refnumber by 1  \count\temptwo=\refnumber \fi
    \tempthr=100 \advance\tempthr by #3
   \ifnum\count\tempthr=0
    \advance\refnumber by 1  \count\tempthr=\refnumber \fi
 \testone=\count\tempone \testtwo=\count\temptwo \testthr=\count\tempthr
 \sortthree\testone\testtwo\testthr
   \test=\testthr  \advance\test by -2
 \ifnum\test=\testone    \test=\testtwo  \advance\test by -1
    \ifnum\test=\testone   
    \ [\the\testone--\the\testthr]\fi \advance\temptwo by 1
  \else
     \ [\the\testone,\the\testtwo,\the\testthr]    
 \fi}

\def\reffour#1#2#3#4{\tempone=100 \advance\tempone by #1
   \ifnum\count\tempone=0
    \advance\refnumber by 1  \count\tempone=\refnumber \fi
    \temptwo=100 \advance\temptwo by #2
   \ifnum\count\temptwo=0
    \advance\refnumber by 1  \count\temptwo=\refnumber \fi
    \tempthr=100 \advance\tempthr by #3
   \ifnum\count\tempthr=0
    \advance\refnumber by 1  \count\tempthr=\refnumber \fi
    \tempfor=100 \advance\tempfor by #4
   \ifnum\count\tempfor=0
    \advance\refnumber by 1  \count\tempfor=\refnumber \fi
 \testone=\count\tempone \testtwo=\count\temptwo \testthr=\count\tempthr
 \testfor=\count\tempfor
 \sortfour\testone\testtwo\testthr\testfor
   \test=\testthr \advance\test by -1
   \ifnum\testtwo=\test   \test=\testtwo \advance\test by -1
    \ifnum\testone=\test  \test=\testfor \advance\test by -3
     \ifnum\testone=\test \ [\the\testone--\the\testfor]
     \else \ [\the\testone--\the\testthr,\the\testfor]
     \fi
    \else  \test=\testfor \advance\test by -1
     \ifnum\testthr=\test \ [\the\testone,\the\testtwo--\the\testfor]
     \else\ [\the\testone,\the\testtwo,\the\testthr,\the\testfor]
     \fi
    \fi
   \else \ [\the\testone,\the\testtwo,\the\testthr,\the\testfor]
   \fi}

\def\reffive#1#2#3#4#5{\tempone=100 \advance\tempone by #1
   \ifnum\count\tempone=0
    \advance\refnumber by 1  \count\tempone=\refnumber \fi
    \temptwo=100 \advance\temptwo by #2
   \ifnum\count\temptwo=0
    \advance\refnumber by 1  \count\temptwo=\refnumber \fi
    \tempthr=100 \advance\tempthr by #3
   \ifnum\count\tempthr=0
    \advance\refnumber by 1  \count\tempthr=\refnumber \fi
    \tempfor=100 \advance\tempfor by #4
   \ifnum\count\tempfor=0
    \advance\refnumber by 1  \count\tempfor=\refnumber \fi
    \tempfiv=100 \advance\tempfiv by #5
   \ifnum\count\tempfiv=0
    \advance\refnumber by 1  \count\tempfiv=\refnumber \fi
 \testone=\count\tempone \testtwo=\count\temptwo \testthr=\count\tempthr
 \testfor=\count\tempfor \testfiv=\count\tempfiv
 \sortfive\testone\testtwo\testthr\testfor\testfiv
  \test=\testthr \advance\test by -1
  \ifnum\testtwo=\test   \test=\testtwo \advance\test by -1
   \ifnum\testone=\test  \test=\testfor \advance\test by -3
    \ifnum\testone=\test \test=\testfiv \advance\test by -4
     \ifnum\testone=\test\ [\the\testone--\the\testfiv]
     \else\ [\the\testone--\the\testfor,\the\testfiv]
     \fi
    \else \ [\the\testone--\the\testthr,\the\testfor,\the\testfiv]
    \fi
   \else  \test=\testfor \advance\test by -1
    \ifnum\testthr=\test \test=\testfiv \advance\test by -2
     \ifnum\testthr=\test \ [\the\testone,\the\testtwo--\the\testfiv]
     \else \ [\the\testone,\the\testtwo--\the\testfor,\the\testfiv]
     \fi
    \else\ [\the\testone,\the\testtwo,\the\testthr,\the\testfor,\the\testfiv]
    \fi
   \fi
  \else \test=\testfor \advance\test by -1
   \ifnum\testthr=\test \test=\testfiv \advance\test by -2
    \ifnum\testthr=\test\
[\the\testone,\the\testtwo,\the\testthr--\the\testfiv]
    \else\ [\the\testone,\the\testtwo,\the\testthr,\the\testfor,\the\testfiv]
    \fi
   \else\ [\the\testone,\the\testtwo,\the\testthr,\the\testfor,\the\testfiv]
   \fi
  \fi}

\def\refitem#1#2{\temp=#1 \advance \temp by 100 \setbox\count\temp=\hbox{#2}}

\def\sortfive#1#2#3#4#5{\sortfour#1#2#3#4\relax
   \ifnum#5<#4\relax \test=#5\relax #5=#4\relax
     \ifnum\test<#3\relax #4=#3\relax
       \ifnum\test<#2\relax #3=#2\relax
         \ifnum\test<#1\relax  #2=#1\relax  #1=\test
         \else #2=\test \fi
       \else #3=\test \fi
     \else #4=\test \fi \fi}

\def\sortfour#1#2#3#4{\sortthree#1#2#3\relax
    \ifnum#4<#3\relax \test=#4\relax #4=#3\relax
       \ifnum\test<#2\relax #3=#2\relax
          \ifnum\test<#1\relax #2=#1\relax #1=\test
          \else #2=\test \fi
       \else #3=\test \fi \fi}

\def\sortthree#1#2#3{\sorttwo#1#2\relax
       \ifnum#3<#2\relax \test=#3\relax #3=#2\relax
          \ifnum\test<#1\relax #2=#1\relax #1=\test
          \else #2=\test \fi \fi}

\def\sorttwo#1#2{\ifnum#2<#1\relax \test=#2\relax #2=#1\relax #1=\test \fi}


\def\setref#1{\temp=100 \advance\temp by #1
   \ifnum\count\temp=0
    \advance\refnumber by 1  \count\temp=\refnumber \fi}

\def\printreference{\totalnumber=\refnumber
           \advance\totalnumber by 1
           \itemnumber=0
           \loop \advance\itemnumber by 1  
                 \ifnum\itemnumber<\totalnumber
                 \item{[\the\itemnumber]} \unhbox\itemnumber \repeat}

%
%
%
%
Rapid anomalous baryon number violating `sphaleron' interactions present
at high temperatures in the standard model certainly play an important
role in determining the baryon asymmetry of the universe.  Despite the large
effort invested\ref{1} however, we still lack a convincing theory of
baryogenesis at the electroweak phase transition (EPT).
What is much better established, is that baryon number was badly
broken at temperatures above the EPT\reftwo{2}{3}, which is known to greatly
alter the more conventional pictures of baryon number production, e.g.\ in
GUTs\ref{4}. Moreover, the necessity of sphaleron reprocessing of
particle asymmetries has made it possible to devise new schemes of
baryogenesis\refthree{5}{6}{7}. For example in leptogenesis\ref{5} one first
generates an asymmetry in the lepton number $L$, which then becomes (partly)
reprocessed into a baryon asymmetry by sphalerons.

Loosely speaking the effect of sphaleron interactions is to break $B+L$
while conserving $B-L$. Then the equilibrium conditions due to various
`ordinary' and anomalous interactions imply\reftwo{3}{8}, to
lowest order, that the $B$ and $L$ asymmetries are proportional to $B-L$.
Thus a universe that initially has $B-L=0$, as is the case in the
simplest GUTs, should have zero baryon asymmetry. This realization
added  motivation for studying the generation of a baryon number
at the EPT.  Also, any source of new exotic baryon or lepton number violation
together with the  sphalerons has the potential to destroy the baryon
asymmetry,
even with a nonzero primordial component of $B-L$. To avoid this outcome,
it appeared to be necessary to place stringent limits on various effective
operators\reffour{9}{8}{10}{101}\hskip-1mm. These limits however, were shown
to be weakened
significantly due to the effective conservation of $e_R$-number
at temperatures $T \gsim 10$ TeV, and to only apply
to flavour independent couplings\reffour{7}{10}{11}{13}\hskip-1mm.

In more accurate treatment of the equilibrium conditions\reftwo{13}{12},
it was shown that even in
the case $B-L=0$ -- assuming that the EPT is weakly enough first order --
the vacuum mass effects in the broken phase (re)generate a small but
nonvanishing  equilibrium value for $B$.  Still it was
maintained that above the EPT temperature $T_C$, $B$ should be zero.
In that case, should the EPT be of first order, as might yet be the
case even in the minimal standard model\ref{14}, the baryon number
would again be zero. (Unless of course, it  was generated
{\it at} the phase transition)

In this letter we will show that the baryon and lepton numbers remain
nonvanishing {\it throughout} the evolution of the universe, even in
the case $B-L=0$, if there are lepton flavour
asymmetries. Above $T_C$ a small nonzero $B$ is preserved due to the
nonidentical dispersion  relations of particles in a plasma.
We will first find the effective thermal
masses of particles in a primeval plasma, and then show how their
presence alters the chemical equilibrium conditions so as to give rise
to a nonzero $B$. We discuss the implications of our results
to the Affleck-Dine\ref{6} type baryogenesis,
where large initial asymmetries may be generated\ref{15}.
\jump
When a particle propagates in a plasma, it continuously interacts with
other particles in the background. Due to these interactions, for example
the Dirac equation for a massless fermion becomes
$$
[(1+a)\Slash k + b\Slash u]\Psi = 0,
\eqno{(1)}
$$
where $a$ and $b$ are some perturbatively calculable momentum and energy
dependent functions\ref{16} that vanish at the zero temperature limit, and
$u$ is the four velocity of the plasma.
The operator multiplying $\Psi$ in (1) is just the inverse propagator, the
zeros of which define the dispersion relation $E = E(k)$. The functional
form of the dispersion relation is in general rather complicated, but
one can introduce effective `thermal' masses $M(T)$ such that to
very good accuracy\ref{16}
$$
E(k) \simeq  \sqrt {k^2+M^2(T)}.
\eqno{(2)}
$$
Thermal masses get contributions from both gauge and Higgs
interactions and in general are different for different
chiralities and families.  Defining $x_i \equiv M_i(T)/T$,
one finds the following expressions in the standard model:
$$
\eqalign{
x_{\ell_i L}^2 &= {3 \over 32}g^2+{1\over 32}g'^2 + {1\over 16} h^{e 2}_i,
 \cr
x_{e_i R}^2 &= {1 \over 8}g'^2 + {1\over 8} h^{e 2}_i,
 \cr
x_{q_j L}^2 &= {3\over 32}g^2 + {1\over 288}g'^2 +
{1\over 6} g_s^2 +{1\over 16} (h_j^{u 2} +h_j^{d 2}) \cr
x_{u_j R}^2 &= {1 \over 18}g'^2 + {1\over 6} g_s^2+ {1\over 8} h_j^{u 2} \cr
x_{d_j R}^2 &= {1 \over 72}g'^2 + {1\over 6} g_s^2+ {1\over 8} h_j^{d 2}, \cr}
\eqno{(3)}
$$
where $g, g'$ and $g_s$ are the usual $SU(2), U(1)_Y$ and $SU(3)$
couplings and the Yukawa couplings $h^f$ derive from the lagrangean
${\cal L}_Y =\sum_{ij} ( h^e_{ij} \bar{L_i} \Phi e_{jR} + h^u_{ij}
\bar{Q_i} \tilde{\Phi} u_{jR} +  h^d_{ij} \bar{Q_i} \Phi d_{jR}) + h.c.$,
with $\tilde{\Phi} = i\sigma_2\Phi$. We choose the  RH fermion
basis such that $h^{f \dagger} h^f$ is diagonal,  the LH
lepton basis such that $h^e h^{e \dagger}$ is diagonal, and the
LH quarks such that $h^u h^{u \dagger} + h^d h^{d \dagger}$
is diagonal. In this basis the eigenvalues of $h^{\dagger} h$
and $h h^{\dagger}$ are the same, and equal to $2 m_f^2/v^2$,
where $v$ is the Higgs vacuum expectation value (= 246 GeV).
Note that we do not know the eigenvalues of the matrix
$h^u h^{u \dagger} + h^d h^{d \dagger}$. However, since
the CKM mixing angles are small, we can assume that
the eigenvalues $\sim m^2_{u_i} + m^2_{d_i}$. In any case,
this is not important for our calculation.
 Numerically one finds\footnote{$^1$}{\fnstyle We assumed that the
coupling constants take their values measured at the weak scale.
More accurately we could write, e.g.\ $x_{tL}^2 \simeq 0.23\chi
+ 0.10$, with $\chi = \ln(M_Z/\Lambda_{QCD})/\ln(T/\Lambda_{QCD}).$}
for example $x_{\ell,L} \simeq 0.21$, $x_{\ell,R} \simeq 0.13$
and $x_{tL} \simeq 0.58$ (see also table 1 below).

Dispersion relations for bosons can be similarly derived and approximated
by thermal masses. The thermal mass of the Higgs boson turns out to
give $x_h \simeq 0.59$, assuming $T\gg T_C$, $m_H = 60$GeV and
$m_{\rm top}= 174$GeV;
for  a complete expression see e.g.\ ref.\ref{11}, but the actual value
of $x_h$ is not important here.  The thermal masses of gauge bosons will
turn out to be irrelevant for our purposes.

Thermal masses can have an effect on baryon number through the boundary
conditions, such as charge neutrality $Q_{em}=0$ (cf.\ equation (7) below).
Total charges are proportional to asymmetries in particle densities,
whereas the equilibrium conditions are relations between chemical
potentials. It is then clear that finite mass effects can modify the
solutions to equilibrium equations subject to boundary conditions
such that nontrivial solutions may exist\ref{12}.

Indeed, the occupation number of a given particle species with momentum
$k$ (in the rest frame of the heat bath) is given by
$$
n(k,T,\mu_i) =  g (\exp((E(k)-\mu_i)/T) \pm 1)^{-1},
\eqno{(4)}
$$
where +(-) refers to fermions (bosons), and $g$ is the
number of internal degrees of freedom of the particle. Then the asymmetry
in that particle species is
$$
\eqalign{
\Delta n_i  &= \int {{\rm d}^3k\over (2\pi )^3}(n(k,T,\mu_i)-n(k,T,-\mu_i)) \cr
&\equiv {g \over 6}({\mu_i T^2})\times (a_\pm-\delta_i),\cr}
\eqno{(5)}
$$
where we have expanded to first order in ${\mu \over T}$,
and $a_+ = 1$ and $a_- = 2$. Assuming that
the dispersion relation is of the form (2), the function $\delta$  becomes
(In the notation of ref.\ref{13} $\delta = a_\pm - \alpha$.)
$$
\delta_\pm =  a_\pm - {6\over \pi^2} \int_x^\infty \dr y \; y
\sqrt{y^2-x^2} {e^y \over (e^y \pm 1)^2}
\eqno{(6)}
$$
For fermions one finds to first order
$\delta_+ \simeq {3\over 2\pi^2}x^2$, which is a very good
approximation for all quarks and leptons. In the bosonic case
the integral expression for $\delta_-$ is nonanalytic near $x=0$,
but one can show that for $x\lsim1$ to good accuracy $\delta_-
\simeq x$.

$$
\vbox{\tt \settabs 5\columns \offinterlineskip
\hrule \vskip2pt \hrule \vskip 7pt
\+  \rm Particle  & $x$ & $\delta$ \cr
\vskip 6pt \hrule \smallskip
\+  $t_L\;(b_L)$& $0.58$ & $4.9\times 10^{-2}$ \cr \vskip2pt
\+  \rm other $q_L$& $0.52$ &  $4.0\times 10^{-2}$\cr \vskip2pt
\+  $t_R$& $0.60$ &  $5.3\times 10^{-2}$\cr\vskip2pt
\+  \rm other $q_{u_R}$& $0.49$ &  $3.6\times 10^{-2}$\cr \vskip2pt
\+  \rm $q_{d_R}$& $0.48$ &  $3.4\times 10^{-2}$\cr \vskip2pt
\+  $i_L$& $0.21$ &  $6.7\times 10^{-3}$\cr \vskip2pt
\+  $\ell_R$& $0.13$ &  $2.4\times 10^{-3}$\cr \vskip2pt
\+  $H$& $0.59$ &  $0.51$\cr
\vskip 4pt \hrule \vskip2pt \hrule
}$$
{\figuref {\bf Table 1.} Shown are the effective masses $x\equiv M(T)/T$ and
corresponding values of $\delta$-functions as given by equation (6). We took
the QCD coupling corresponding to the scale $M_Z$ (see footnote 1). Only the
top quark Yukawa coupling is large enough to show up in these results. We have
used $m_{\rm top} = 174 GeV$ and $\alpha_s = 0.11$.}
\jump
We will now solve for the baryon number in the presence of finite
temperature mass effects in a universe with $B-L=0$, but $L_i \neq 0$.
At $T \gsim T_C$, there are
additional boundary conditions of vanishing total charge-
and isospin densities, so that the complete set of
boundary conditions to be used is:
$$
Q_{em}  = 0 \qquad \qquad Q_3  = 0 \qquad \qquad B-L  = 0.
\eqno{(7)}
$$
Rapid decays and inverse decays (when kinematically allowed) and various
$2-2$ scattering processes induce the usual set\footnote{$^2$}{We can
assume that the temperature is less than ${\cal O}(few)$TeV, so that
also right handed electrons are in thermal equilibrium\ref{11}.}
of `ordinary' equilibrium relations between chemical potentials\ref{8}:
$$
\eqalign{
\mu_- + \mu_0 = \phantom{'}\mu_W &\phantom{hann:}\hskip-1pt
\mu_{u_R} - \mu_{u_L} =\phantom{-}\mu_0 \cr
\mu_{d_R} - \mu_{d_L} \hskip 1pt=-\mu_0 &\phantom{hanna}\hskip 3pt
\mu_{i_R}-\mu_{i_L} = -\mu_0 \cr
\mu_{d_L} - \mu_{u_L} =\phantom{:}\mu_W &\phantom{hannam}
\mu_{i_L} - \mu_i =  \mu_W, \cr}
\eqno{(8)}
$$
where the subscipts indicate particle species --- ``0" and $``-"$
being the  Higgs doublet, $i$ a neutrino species, and $i_L$ and
$i_R$ are charged leptons of the $i$th generation. Additionally,
there is an equilibrium condition due to anomalous sphaleron
processes which, with help of (8), can be written as
$$
9\mu_{u_L} + 6\mu_W + \mu = 0,
\eqno{(9)}
$$
where $\mu \equiv \sum_i \mu_i$.
We can use the equations (5) for particle asymmetries along with the
equilibrium conditions (8) to write the global charges as
$$
\eqalign{
Q \; & \; =\; (6-2\Delta_u+\Delta_d)\; \mu_{uL}
- (18 - 4\delta_W - 2\delta_- - \Delta_d  -\Delta_e)\; \mu_W \cr
&\phantom{hai}+ (14-2\Delta_{uR}
-\Delta_{dR} -\Delta_{eR} - 2\delta_-)
\; \mu_0 - 2\mu + \Delta\mu_L + \Delta\mu_R \cr
Q_3 \; & \; =\; -(11 - 4\delta_W -\delta_- -{3\over 2}\Delta_{dL}
- {1\over 2}\Delta_{eL})\; \mu_W \cr
&\phantom{hai}+ {3\over 2}(\Delta_{dL} - \Delta_{uL}) \; \mu_{uL}
+ {1\over 2}(\Delta\mu_{L}- \Delta\mu_{\nu}) + \mu_0 (\delta_0 - \delta_-)\cr
B \; & \; =\;  (12 - \Delta_q)\; \mu_{uL} +
(6 - \Delta_d) \;\mu_W +
(\Delta_{dR} - \Delta_{uR})\; \mu_0\cr
L \; & \; =\; 3\mu - \Delta\mu_{\nu} -\Delta\mu_L -\Delta \mu_R
+ (6-\Delta_{e})\;\mu_W
- (3- \Delta_{eR})\;\mu_0, \cr}
\eqno{(10)}
$$

\noindent where we have dropped the common multiplicative factor
${15\over 4\pi^2g_* T}$, (we define $L \equiv \Delta n/s$, where $s$
is the entropy density so that $g_*$ is the number of relativistic
degrees of freedom taken to be 106.75 from now on) and defined

$$
\eqalign{
\Delta\mu_{\nu} &\equiv \sum_i \mu_i\delta_{i} \cr
\Delta\mu_X &\equiv \sum_i \mu_i\delta_{iX} \qquad  X = L, R\cr
\Delta_{fX} &\equiv \sum_i\delta_{f_{iX}} \qquad X= L,R ~~ f_i = u_i,
d_i, e_i \cr}
\eqno{(11)}
$$
and moreover $\Delta_f \equiv \Delta_{fL}+\Delta_{fR}$ and
$\Delta_q \equiv \Delta_{u}+\Delta_{d}$.  Notice that the chemical
potential $\mu_i$ in definitions (11) is always that of neutrinos.
We must now apply the boundary conditions (7) to the expressions (10).
At first sight this looks rather messy, but fortunately some of the
coefficients in (10) turn out to be zero, so that an analytic
solution of decent length can be obtained for $B$.

First of all, dispersion relations of particles within the same weak
doublet are identical, so that, $\delta_{f_{uL}}-\delta_{f_{dL}}= 0$,
$\delta_{0}-\delta_{-}= 0$, and
$\Delta \mu_{L}  = \Delta \mu_{\nu}$.
Then only the term proportional to $\mu_W$ remains in the expression
for $Q_3$ and therefore, even with thermal corrections we get $\mu_W=0$.
After this simplification it is rather straightforward to
proceed: we use (9) to solve $\mu_{uL} = -\mu/9$ and then solve $\mu_0$
from the charge equation:
$$
\mu_0 = {1 \over 14 - \Delta_1} \; \left\{
({8 \over 3} - \Delta_2) \mu  - \Delta\mu_L -\Delta\mu_R \right\},
\eqno{(12)}
$$
where we used the shorthand notations $\Delta_1 \equiv 2\Delta_{uR}
+ \Delta_{dR} + \Delta_{eL} + 2\delta_{ 0}$ and
$\Delta_2 \equiv {1\over 9}(2\Delta_u - \Delta_d)$.
Finally, the last constraint $B-L=0$ becomes a consistency
condition for $\mu$:
$$
\left\{ {13\over 3} - {\Delta_q\over 9} - ({8\over 3}-\Delta_2)
{3+\Delta_3 \over 14 - \Delta_1}\right\}\; \mu
= \Delta\mu_L + (1-{3+\Delta_3 \over 14 - \Delta_1})
(\Delta\mu_L +\Delta\mu_R),
\eqno{(13)}
$$
where $\Delta_3 \equiv \Delta_{dR} - \Delta_{uR} - \Delta_{eR}$.
If all thermal corrections vanish, equation (13) has only the trivial
solution $\mu=0$, which would immediately lead to $B=L=0$, as
indeed was earlier thought to be the case\refthree{8}{12}{13}.

At first sight, (13) might be thought to be giving rise to relatively
large asymmetries, because the gauge contributions to effective masses
are not particularily small. However, the gauge contributions are the
{\it same} for all families and their added contribution is
proportional to $\mu$. Indeed, we can write the $\Delta\mu_X$-terms
in the R.H.S\ of equation (13) as
$$
\Delta\mu_X = \delta^{g}_X\; \mu + \sum_i \delta^Y_{iX} \mu_i,
\qquad X = L, R
\eqno{(14)}
$$
where $ \delta^{g}$ and $ \delta^Y$ respectively are the gauge
and Yukawa mass corrections (see equation (3)).
Thus the equation (13) is of the form $A\mu = a\mu + c$, where
$c$ is nonzero, and hence a nontrivial solution exists, only because
lepton family number is conserved and
the lepton Yukawa interactions differ from one family to another.
This is in close analogy to the situation when
$T\lsim T_C$\reftwo{12}{13}. Using (14) one can rewrite (13) as
$$
\eqalign{
\left\{ {13\over 3} - {\Delta_q\over 9} - ({8\over 3}-\Delta_2)
{3+\Delta_3 \over 14 - \Delta_1}
-\delta^g_L \right. & \left. \hskip-0.7mm -\;
(1-{3+\Delta_3 \over 14 - \Delta_1})
(\delta^g_L+\delta^g_R)  \right\}  \mu \cr
&= (4-{3(3+\Delta_3) \over 14 - \Delta_1}) \Delta\mu_L^Y, \cr}
\eqno{(15)}
$$
which shows explicitly the dependence on Yukawa terms
$\Delta\mu_L^Y \equiv \sum_i \delta_{i_L}^Y\mu_i$. In principle
it is easy to trace  backwards the steps from (15) to (12) to (10) and
solve for $B$ exactly. However, glancing at the size of the various
thermal correction terms (see also table 1),
$$
\Delta_q \simeq 0.49 \quad
\Delta_1 \simeq 1.39 \quad \Delta_2 \simeq 0.03 \quad
\Delta_3 \simeq -0.03 \quad \delta_L^g \simeq 0.007 \quad
\delta_R^g \simeq 0.002,
\eqno{(16)}$$
one can see that all of the quantities in (16) are small enough to
be neglected. Then one finds a simple expression for $B\propto 12\mu_{u_L}
\simeq -{4\over 3}\mu$,
$$
B \simeq - {94 \over 79}({15\over 4\pi^2g_*})({ \Delta\mu^Y_L \over T}).
\eqno{(17)}$$
Including all $\Delta$-terms (with $x_h=0.59$) one obtains $B\simeq
- 4.15\times 10^{-3} \Delta\mu^Y_L/T$, which is less than 3 percent
off from the value given by (17)!
The Yukawa couplings of leptons are rather small and
obey the hierarchy $h_e \ll h_\mu \ll h_\tau$ (see table (2)), so that
to very high accuracy $\Delta\mu^Y_L \simeq \delta_\tau^Y \mu_\tau
\propto \delta_\tau^YL_\tau$.
\jump
$$
\vbox{\tt \settabs 5\columns \offinterlineskip
\hrule \vskip2pt \hrule \vskip 7pt
\+  \rm Particle & $h_\ell$ & $x^2_Y$ & $\delta^Y$ \cr
\vskip 6pt \hrule \smallskip
\+  $\tau_L,(\nu_\tau)$ & $1.03\times 10^{-2}$ & $6.6\times 10^{-6}$ &
$1.0\times 10^{-6}$\cr \vskip2pt
\+  $\mu_L,(\nu_\mu)$ & $6.09\times 10^{-4}$ & $2.3\times 10^{-8}$ &
$3.5\times 10^{-9}$\cr\vskip2pt
\+  $e_L, (\nu_e)$& $2.94\times 10^{-6}$ & $5.4\times 10^{-13}$ &
$8.2\times 10^{-14}$ \cr \vskip2pt
\vskip 4pt \hrule \vskip2pt \hrule
}$$
{\figuref {\bf Table 2.} Shown are the Yukawa contributions to the lepton
thermal masses and the corresponding values of $\delta$ functions. For
right handed charged leptons use $x_R^2 = 2x_L^2$ and $\delta_R
= 2\delta_L$.}
\jump
We still need to express $B$ in terms of primordial quantities.
It is easy to show that if $B-L$ = 0 and ${1\over 3}B - L_i$ are
conserved, then
$$
L_\tau (t) ={1 \over 6} ( [ B + L](t) + 2\Delta L_{\tau e} +
2\Delta L_{\tau \mu}).
\eqno{(18)}
$$
where $L_\tau \equiv L_{\tau_L} +  L_{\tau_R} +  L_{\nu_\tau}$
and $\Delta L_{ij} \equiv L_i-L_j$.
The sphalerons force $B+L$ and the Higgs chemical potential
to be zero, up to the small mass corrections that we are
calculating, so that
$L_{\tau} \simeq {45\over4\pi^2g_*}({\mu_{\tau}\over T})$,
and
$$
B \simeq -{94 \over 79} {3 \over 2 \pi^2} x_{\tau}^2
{ (\Delta L_{\tau e} + \Delta L_{\tau \mu}) \over 9}
\simeq 1.3 \times 10^{-7}
 (\Delta L_{e\tau} + \Delta L_{\mu \tau}),
\eqno{(19)}
$$
This is our final result.
It should be noted that mass effects alone are not enough to
provide a nonzero $B$, although they set the scale of $B$ in terms
of primordial asymmetries. Instead, primordial values of leptonic
asymmetries must be different, i.e. $\Delta L_{ij} \neq 0$!

Let us compare our result (19) to that of refs.\reftwo{12}{13},
where the situation at $T<T_C$ is considered.
Equation (23) for $B$ in ref.\ref{13} is similar to our
equation (17) and it can also be brought into the form (19).
Neglecting all mass corrections except the bare lepton Yukawa terms
responsible for a nonzero result, one obtains from ref.\ref{13}
$B \simeq few \times 10^{-7} (\Delta L_{e \tau} + \Delta L_{\mu\tau})$.
Of course, if the EPT is of second order or
weakly enough first order, it is the latter result that
is relevant.  However, if the EPT is of first order, then
sphaleron processes drop out of equilibrium instantaneously
in the broken phase, the equilibrium conditions below $T_C$ never
get realized and the value of $B$ given by (19) gets frozen
into the system.  The point we wish to make is that {\it either}
way, a nonzero fraction of primordial baryon number survives.

In order for a component of primordial baryon number to survive
at the level of the observed asymmetry $B_{now} = n_B/s \simeq
4\times10^{-11}$\ref{17}, primordial asymmetries must be of
order  $\sim 10^{-4}$. This is perhaps at best marginally
possible in the simplest GUT scenarios, but it is much more
feasible in the Affleck-Dine mechanism\ref{6}, where a baryon
asymmetry is generated
during the decay of a scalar condensate, which can have a
large component of baryon or lepton number stored within.
It was previously found\ref{15} that sufficiently large ($B \gsim
10^{-2}$) primordial $B+L$ asymmetries can survive in the
Affleck-Dine scenario, because the squark condensate does
not evaporate before the electroweak phase transition, giving
the $W$ a large mass and keeping the sphalerons out of
thermal  equilibrium. However, a large amount of
entropy production is then required after the phase transition.
We have shown here that within the Standard Model, given a
slightly smaller ``large'' primordial asymmetry $\sim 10^{-4}$,
it would not be washed out, but rather
be diluted by the sphalerons to give approximately the right
baryon asymmetry today, so that no entropy production
is required.

Let us finally discuss the apparent limitation that nonzero
$\Delta {L_{ij}}'$s are required for the mechanism to work. First
of all, it is not at all unnatural to obtain
differences in the asymmetries that are of the order of the
asymmetries themselves.  On the contrary, in e.g.\  A-D models it
is natural to expect that the scalar condensate initially lies in
some arbitrary direction in the slepton space, which in
general is not symmetric in flavours. In such case the subsequent
evolution and decay of the condensate naturally gives rise to large
differences in lepton asymmetries.  Thus large $\Delta {L_{ij}}'$s
certainly can be produced.  Secondly, in models with $B-L$ symmetry
$\Delta {L_{ij}}'$s are always conserved quantities.
Moreover, even in the case that the lepton number was broken at
some higher temperature scale due to some new exotic interactions\ref{9},
$B$ typically would not vanish. This is due to the approximate $e_R$-number
conservation, which protects the right handed electron asymmetry, and
hence a nonzero $\Delta L_{\tau e}$ at any temperature $T\gsim 10$
TeV\ref{11}. Thus, except in some very speculative cases (for details
see ref.\ref{11}) the survival of a nonzero $B$ is always guaranteed.

\jump
In conclusion, we have computed the equilibrium value of the baryon
asymmetry $B$ in the early universe above the electroweak phase transition
temperature $T_C$. We have found that due to thermal effects (nontrivial
dispersion relations) a small, but nonzero value of $B$ persists throughout
the evolution of the universe, even in the case $B-L=0$.
The final value of $B$ is roughly seven orders of magnitudes smaller than
the primordial value, with the scale set
by the Yukawa coupling of the tau lepton. Our results complement the
analysis of refs.\reftwo{12}{13}, where the {\it vacuum} mass effects at
the broken phase at $T <T_C$ were shown to also secure a nonzero $B$.
Combining these results one has the proof of persistence of a fraction
of primordial baryon asymmetry regardless of the order of the
electroweak phase transition. This result favours
Affleck-Dine type baryogenesis models, which
otherwise could produce too large a final baryon asymmetry;
the scale of relaxation of $B$ due to modified equilibrium processes
can make these theories work even for rather large initial asymmetries.
\jump
\centerline {\bf Acknowledgements}
\jump
K.K.\ wishes to thank Jim Cline and S.D.\ would like to thank H.\ Murayama
and J.\ March-Russel for useful discussions. This research is supported
by the DOE grant DE-FB02-94ER40823, by NSF grant AST 91-20005, by NSERC,
KAO's Presidental Young Investigator Award and by (K.K.) the Finnish
Academy.
\vfill\eject
\centerline {\subsection References}
\jump
\refitem{1} {A.G.\ Cohen, D.B.\ Kaplan and A.E.\ Nelson, Ann.\ Rev.\ of Nucl.\
            Sci.\ {\bf 43} (1993) 27, and references therein.}
\refitem{2} {V.~Kuzmin, V.~Rubakov, and M.~Shaposhnikov, Phys.~Lett.~{\bf B155}
             (1985) 36.}
\refitem{3} {P.\ Arnold and L.\ McLerran, Phys.\ Rev.\ {\bf D37} (1988) 1020.}
\refitem{4} {J.N. Fry, K.A. Olive and M.S. Turner, Phys.\ Rev.\ {\bf D22}
(1980)
             2953.}
\refitem{5} {M.\ Fukugita and T.\ Yanagida, Phys.\ Lett.\ {\bf B174} (1986) 45;
             M.\ Luty, Phys.\ Rev.\ {\bf D45}(1992)455; B.A.\ Campbell,
             S.\ Davidson, and K.A.\ Olive,  Alberta Thy-38-92, CfPA 92-035,
             UMN-TH-1114/92.}
\refitem{6} {I.\ Affleck and M.\ Dine, Nucl.\ Phys.\ {\bf B249} (1985) 361.}
\refitem{7} {B.A.\ Campbell, S.\ Davidson, J.\ Ellis, and K.A.\ Olive,
             Phys.\ Lett.\ {\bf B297} (1992) 118.}
\refitem{8} {J.\ Harvey and M.\ Turner, Phys.\ Rev.\ {\bf D42}(1990)3344.}
\refitem{9} {M.~Fukugita and T.~Yanagida, Phys.~Rev.~{\bf D42} (1990) 1285.}
\refitem{10}{A.E.\ Nelson and S.M.\ Barr Phys.\ Lett.\ {\bf B258} (1991) 45.}
\refitem{101} {W.\ Fishler, G.F.\ Giudice, R.G.\ Leigh and S.\ Paban, Phys.\
             Lett.\ {\bf B258} (1991) 45. B.A.~Campbell, S.~Davidson, J.~Ellis
             and K.A.~Olive, Phys.~Lett.~{\bf B256} (1991) 457; Astroparticle
             Physics {\bf 1} (1992) 77.}
\refitem{13}{H. Dreiner and G. Ross, Nucl.\ Phys.\ {\bf B410} (1993) 188.}
\refitem{11}{J.M.~Cline, K.~Kainulainen and K.A.~Olive, Phys.\ Rev.\ Lett.\
            {\bf 71} (1993) 2372, Phys.\ Rev.\ {\bf D}, in press.}
\refitem{12}{V.~Kuzmin, V.~Rubakov, and M.~Shaposhnikov, Phys.~Lett.~{\bf B191}
             (1987) 171.}
\refitem{14}{K.\ Kajantie, K.\ Rummukainen and M.\ Shaposnikov, Nucl.\ Phys.\
             {\bf B407} (1993) 356, M.\ Shaposnikov, Phys.\ Lett.\ {\bf B316}
             (1993) 112.}
\refitem{15}{S.\ Davidson, H.\ Murayama and K.A.\ Olive, University Minnesota
             preprint, UMN-TH-1240/94 1994.}
\refitem{16}{H.A.\ Weldon Phys.\ Rev.\ {\bf D26} (1982) 2789.}
\refitem{17}{K.A.\ Olive, D.N.\ Schramm, G.\ Steigman and T.P.\ Walker,
             Phys.\ Lett.\ {\bf B236} (1990) 454; T.P.\ Walker, G.\ Steigman,
             D.N.\ Schramm, K.A.\ Olive and H.S.\ Kang, Ap.J.\ {\bf 376}
             (1991) 51.}
\printreference
\jump
\vfill\eject
\null
\nopagenumbers
\vskip -20pt
\vskip 0.2truecm
\line {\hfill hep-ph/yymmddd}
\line {\hfill UMN-TH-1248/94}
\line {\hfill TPI-MINN-94/11-T}
\line {\hfill CfPA-TH-94-19}
\line {\hfill April 1994}
\vskip 2.2truecm
\centerline {\subsection Protecting the Baryon Asymmetry with Thermal Masses}
\vskip 1.5truecm
\centerline {S. Davidson$^{1}$, K.\ Kainulainen$^{2}$ and K.A.\ Olive$^{2}$}
\vskip 0.4truecm
\centerline {$^1$ Center for Particle  Astrophysics, Unviersity of California,}
\centerline {Berkeley, California, 94720}
\centerline {$^{2}$School of Physics and Astronomy, University of Minneapolis}
\centerline {Minneapolis, MN 55455}

\vskip 2.0truecm
\centerline {\bf Abstract}
\vskip 1truecm
\hskip -15pt
We consider the evolution of baryon number $B$ in the early universe under
the influence of rapid sphaleron interactions and show that $B$ will remain
nonzero at all times even in the case of $B-L = 0$. This result arises
due to thermal Yukawa interactions that cause nonidentical dispersion
relations (thermal masses) for different lepton families. We point
out the relevance of our result to the Affleck-Dine type baryogenesis.
\vfill\eject
\end